\documentclass[%
reprint,
amsmath,amssymb,
aps,
pre,
]{revtex4-2}

\usepackage{graphicx}
\usepackage{dcolumn}
\usepackage{bm}
\usepackage[colorlinks,urlcolor=blue,linkcolor=blue,anchorcolor=blue,citecolor=blue]{hyperref} 
\usepackage{amsmath}
\usepackage{subfigure}
\usepackage{color}    
\usepackage[normalem]{ulem}    
\usepackage{hyperref}
\usepackage{enumerate}

\newcommand{\mapN}{\includegraphics[height=1em]{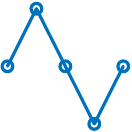}}
\newcommand{\mapNN}{\includegraphics[height=1em]{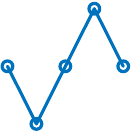}}
\newcommand{\systemN}{\includegraphics[height=1em]{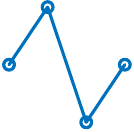}}
\newcommand{\systemNN}{\includegraphics[height=1em]{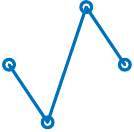}}

\usepackage[utf8]{inputenc} 
\DeclareUnicodeCharacter{200B}{} 

\hypersetup{
	colorlinks=true,       
	linkcolor=blue,        
	citecolor=blue,         
	filecolor=magenta,     
	urlcolor=blue,         
}

\usepackage[most]{tcolorbox}
\usetikzlibrary{decorations.pathmorphing}

\newtcolorbox{deletedsection}{
	blanker, 
	breakable, 
	overlay={\draw[red, thick] (frame.north west) -- (frame.south east); 
		\draw[red, thick] (frame.north east) -- (frame.south west); 
	}
}

\begin{document}
	
	\preprint{APS/123-QED}
	
	\title{​​​​Ultraslow Diffusion Emerging from Local Symmetry Breaking in Quenched Disorder​​​​}
	
	\author{Jin Liu$^{1}$}
	
	\author{Rainer Klages$^{2,3}$}
		
	\author{Huihai Wang$^{4}$}
	
	\author{Kehui Sun$^{1,}$}
	\email{kehui@csu.edu.cn}
	
	\affiliation{$^{1}$ School of Physics, Central South University, Changsha 410083, China}
	\affiliation{$^{2}$ Centre for Complex Systems, School of Mathematical Sciences, Queen Mary University of London, Mile End Road, London E1 4NS, United Kingdom}
	\affiliation{$^{3}$ London Mathematical Laboratory, 8 Margravine Gardens, London W6 8RH, United Kingdom}
	\affiliation{$^{4}$ School of Electronic Information, Central South University, Changsha 410083, China}

	\date{\today}

	
	\begin{abstract}
		Many dynamical systems operate in disordered environments. In this Letter, we investigate how quenched randomness modifies particle diffusion. We show that, in finite-strength potential landscapes, low disorder densities can induce logarithmically ultraslow diffusion. The underlying mechanism originates from local symmetry breaking in particular potential configurations: barrier--well structures form directional channels, causing particles to undergo hierarchical capture, escape, and recapture processes. A coarse-grained channel theory yields an approximate scaling prediction for the mean-square displacement. Counterintuitively, when the disorder density reaches the fully disordered limit, the system returns to normal diffusion, revealing a nontrivial, nonmonotonic relationship between disorder density and transport type. This mechanism is observed in both nonlinear maps and stochastic systems.
	\end{abstract}
	
	\maketitle
	
	Imagine embarking on a long-distance journey where, upon reaching your destination, you engage in a series of localized activities (like lodging and sightseeing) that may trigger smaller-scale movements (e.g., commuting between hotels and restaurants). Remarkably, despite continuous progression of activities, macroscopic analysis reveals that the spatial expansion of your trajectory exhibits gradual deceleration in displacement over time. This phenomenon aligns with the experimental findings of Song $et$ $al.$ \cite{song2010modelling}, demonstrating that the mean square displacement (MSD) in long-term human mobility patterns follows an ultraslow diffusion characterized by $\operatorname{MSD}(t) \sim {{\left( \ln t \right)}^{\alpha }}$. This law is found in a wealth of systems, including relaxation processes in Anderson insulators \cite{vaknin2000aging}, local structural relaxation in DNA \cite{brauns2002complex}, paper crumpling under heavy pistons \cite{matan2002crumpling}, frictional strengthening \cite{bendavid2010slipstick}, grain compactification \cite{richard2005slow}, glass systems \cite{boettcher2011ageing}, and fractional dynamics in languages \cite{watanabe2018empirical}.

	Currently, theoretical studies on ultraslow diffusion remain relatively scarce and are predominantly modeled through stochastic systems. The mechanisms involve the introduction of complex spatiotemporal structures, exemplified by distributed-order fractional derivative models \cite{eab2011fractional,kochubei2008distributed}, structural derivative models \cite{liang2018nonlocal,liang2018diffusion}, and comb models \cite{sandev2016comb,havlin1987anomalous,iomin2016ultraslow}, as well as super-heavy-tailed waiting time distributions as implemented in enhanced continuous-time random walk models \cite{havlin1990longtailed,meerschaert2006stochastic,denisov2013limiting,denisov2011asymptotic}. Alternatively, one-dimensional maps represent the simplest dynamical systems capable of generating diffusion \cite{geisel1982onset,schell1982diffusive,fujisaka1982chaos,geisel1984anomalous,geisel1985accelerated,zumofen1993scaleinvariant,barkai1997crossover,klafter1996beyond,klages1997dynamical,klages2002suppression,barkai2003aging,korabel2007fractal}, when defined on infinite domains, typically by periodic continuation. These maps emerged as simplified representations to capture essential characteristics of driven damped particle motion in a periodic potential \cite{geisel1982onset,schell1982diffusive,fujisaka1982chaos}. While time-dependent periodic potentials represent one limiting case of this scenario, the opposite extreme---disordered potentials---remains understudied in chaotic dynamical systems theory \cite{sarkar2023quench,liu2024quenched}. Across multiple physical domains, it is well-established that static randomness can profoundly alter macroscopic quantities \cite{havlin1987diffusion,zwanzig1988diffusion,popescu1998quenched}. Nevertheless, developing deterministic map versions capable of producing ultraslow diffusion remains highly challenging. To date, meaningful progress has been achieved only in studies by Radons \cite{radons1996suppression,radons2004anomalous} and Dräger and Klafter \cite{drager2000strong}. The former demonstrated suppressed diffusion equivalent to the Sinai model \cite{sinai1982limit}, requiring the overcoming of increasingly deeper Brownian path random potentials in the continuous limit. The latter revealed through a fully deterministic one-dimensional lifted map that ultraslow diffusion originates from the anomalously broad distribution of waiting times in the corresponding random walk process, which under specific conditions can reproduce Sinai's classical results.
	
	In this Letter, we demonstrate a universal mechanism generating ultraslow diffusion. By introducing quenched disordered finite-strength potentials in our models (both deterministic and stochastic versions)---fundamentally different from the deep potential well mechanism underlying Golosov localization \cite{golosov1984localization,bouchaud1990anomalous} in Sinai's and Radons's works---we observe an unexpected nontrivial nonmonotonic transport transition where particle diffusion in low-disorder-density (sparse) potentials (corresponding to ultraslow diffusion) paradoxically becomes slower than that in the fully disordered case (normal diffusion) at extended timescales. We establish that this phenomenon originates from symmetry breaking in localized potential configurations.
	
	\begin{figure}[t]        
		\centering
		\includegraphics[width=0.9\linewidth]{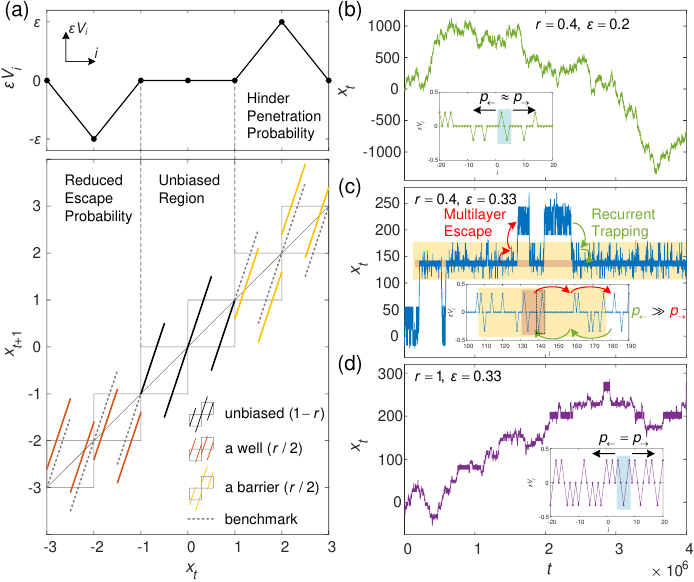}
		\subfigure{\label{fig:1a}}
		\subfigure{\label{fig:1b}}
		\subfigure{\label{fig:1c}}
		\subfigure{\label{fig:1d}}
		\caption{Schematic illustration of the chaotic map modeling under quenched disorder and its representative particle trajectories. (a) The potential landscape comprises static random barriers and wells with finite-strength $\varepsilon$ (upper panel). The wells and barriers in map (\ref{eq:2}) are modeled by vertically shifting the function profiles within adjacent lattice sites (lower panel), thereby modifying particle escape probabilities in corresponding regions. (b) Particle trajectories exhibit Brownian-like motion \cite{dasilva2019correlated} at $r=0.4$ with $\varepsilon=0.2$. (c) When $\varepsilon$ increases to $0.33$ while maintaining $r=0.4$, the particle motion demonstrates distinct transport transition characteristics, where hierarchical escape events and recurrent trapping significantly suppress diffusion efficiency. (d) Brownian-like dynamics is globally restored at long timescales when the disorder density reaches $r=1$ (fully disordered). Insets in (b)--(d) display characteristic configurations of the corresponding potential distributions $\varepsilon V_i$.}
		\label{fig:1}
	\end{figure}

	A piecewise linear lifted map capable of generating normal diffusion \cite{klages1995simple,gaspard1998chaotic,klages1999simple,sato2019anomalous} is defined as
	\begin{equation}
		\label{eq:1}
		\begin{matrix}
			{{M}_{a}}({{x}_{t}})=\left\{ \begin{array}{*{35}{l}}
				a{x_t}, & 0\le {x_t}<\frac{1}{2},  \\
				a{x_t}+1-a, & \frac{1}{2}\le {x_t}<1,  \\
			\end{array} \right. & a>0,  \\
		\end{matrix}
	\end{equation}
	extended to the real domain $\mathbb{R}$ through the periodic condition $M_a(x_t+1)=M_a(x_t)+1$. In our fundamental setup [$a=3$, black solid and dashed lines in Fig. \ref{fig:1a} lower panel], we implement a quenched disorder potential by vertically shifting adjacent lattice functions: Specifically, shifting the left one upward and the right downward by $\varepsilon$ (defined as the potential strength) reduces the particle escape probability to simulate wells [maroon solid line in Fig. \ref{fig:1a}]. Conversely, shifting the left one downward and the right upward by $\varepsilon$ hinders the penetration probability to model barriers of equal strength \cite{felderhof2008escape} (yellow solid line). Each well or barrier occurs with a probability $r/2$. Defining this system as map $Q_1$, its dynamical equation can be expressed as
	\begin{equation}
		\label{eq:2}
		\begin{matrix}
			Q_1: & {{x}_{t+1}}={{M}_{3}}({{x}_{t}})-\varepsilon \Delta {V_i}, & \text{where} & {{V}_{i}}={{\xi }_{i}}{{\mu }_{i}}  \\
		\end{matrix}.
	\end{equation}

	

	This is the central model discussed in our Letter, where $i = [x_t]$ denotes the lattice site index, and $\Delta {{V}_{i}}=V_{i+1}-V_{i}$ represents the potential gradient within the $i$-th lattice site. The disorder density is governed by Bernoulli random variables ${{\xi }_{i}} \sim \mathcal{B}(r)$: When $\xi_i=1$ (occurring with probability $r$), $\mu_i \sim \mathcal{U}\{-1,1\}$ is uniformly sampled, where $-1$ corresponds to wells and $1$ to barriers. Since every two consecutive lattice sites form a single potential unit, this sampling occurs exclusively at even-indexed sites, i.e., $V_i​=0$ for odd $i$. This ensures uniform well/barrier strength. When $\xi_i=0$ (with $1-r$), the $V_i$ remains zero. Our map $Q_1$ maintains chaotic characteristics with Lyapunov exponent $\ln 3$ \cite{sato2019anomalous,klages2007microscopic}.
	
	In a system composed exclusively of either wells or barriers ($\mu_i = -1$ or $1$), particle diffusion is hindered by such structures. Owing to renormalization effects \cite{degennes1975brownian,bassler1987viscous,zwanzig1988diffusion}, the system exhibits normal diffusion with a reduced diffusion coefficient [see Fig. S1(b) of the Supplemental Material (SM) \cite{supplementalMaterial}\nocite{klafter2011first,norris1998markov,lomholt2013microscopic,li2025route,dentz2016selfaveraging,huberman1980noise,dhumieres1982chaotic}]. Intuitively, there is no reason to expect different dynamical transitions in the map $Q_1$ ($\mu_{i} \in \{-1,1\}$). However, simulations demonstrate that when $r=0.4$ and $\varepsilon = 0.2$ (significantly below the localization threshold $\varepsilon _{\mathrm c} =1/3$, as demonstrated in Sec. I of the SM), particles maintain Brownian-like motion [Fig. \ref{fig:1b}]. Conversely, when $\varepsilon = 0.33$ (approaches the critical value) with unchanged $r=0.4$ (i.e., spatially sporadic distribution of wells and barriers), trajectories exhibit prolonged confinement within specific hierarchical layers, where escape events become exceedingly rare and are accompanied by high-probability backtracking events, resulting in significant diffusion retardation [Fig. \ref{fig:1c}]. More intriguingly, under fixed $\varepsilon =0.33$, the fully disordered system ($r=1$, where wells and barriers are ubiquitously distributed) recovers Brownian-like motion at longer timescales [Fig. \ref{fig:1d}], further suggesting a potential nontrivial transport transition implied in Fig. \ref{fig:1c}.

	\begin{figure}[t]       
		\centering
		\includegraphics[width=0.45\linewidth]{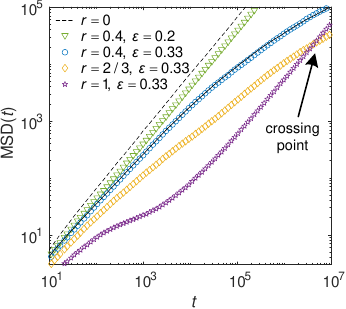}
		\caption{MSD and nonmonotonic transport transition in map $Q_1$. When the potential strength is far from the critical value, $Q_1$ exhibits normal diffusion. In contrast, under near-critical sparse disorder, the growth of the MSD is strongly suppressed and displays ultraslow diffusion. Counterintuitively, in the fully disordered limit $r = 1$, the system recovers normal diffusion at long timescales. The black solid line shows the result for system $Q_2$ at $r=0.4$, $c=5.5$, and $e=0.495$. After rescaling its MSD by a factor of 2, it exhibits the same statistical trend as $Q_1$.}
		\label{fig:2}
	\end{figure}

	This perspective is validated by MSD analysis [Fig. \ref{fig:2}], based on numerical simulations of $1 \times 10^6$ initial points uniformly distributed in $[-1,1]$. When $\varepsilon$ deviates significantly from the critical value, map $Q_1$ exhibits normal diffusion with a reduced diffusion coefficient (green triangles). Under near-critical sparse disorder, however, the map undergoes a subtle dynamical transition: the growth of the MSD is strongly suppressed, giving rise to ultraslow diffusion (blue circles and yellow diamonds). Remarkably, normal diffusion is eventually restored in the fully disordered case $r=1$ (purple stars). At sufficiently long times, the MSD for $r=1$ exceeds that observed for $r<1$, consistent with the observations in Figs. \ref{fig:1b}--\ref{fig:1d}: an intriguing nontrivial transition emerges only in sparse disordered potentials.

	To investigate the origin of this anomalous diffusion, we draw an analogy with the renowned Sinai model \cite{sinai1982limit}. For the lattice version of the Sinai model on infinite integer lattices \cite{chave1999statistical,padash2022local}, formulated as
	\begin{equation}
		\label{eq:3}
		\left\{ \begin{array}{*{35}{l}}
			{{q}_{i,i-1}}=\frac{1+e\left( {{U}_{i}}-{{U}_{i-1}} \right)}{2+c},  \\
			{{q}_{i,i+1}}=\frac{1+e\left( {{U}_{i}}-{{U}_{i+1}} \right)}{2+c},  \\
			{{q}_{i,i}}=\frac{c+e\left( {{U}_{i+1}}+{{U}_{i-1}}-2{{U}_{i}} \right)}{2+c},  \\
		\end{array} \right.
	\end{equation}
	with
	\begin{equation}
		\label{eq:4}
			{{U}_{i}}={{U}_{i-1}}+{{\xi }_{i}} {{\mu }_{i}},
	\end{equation}
	where ${{\xi }_{i}} \sim \mathcal{B}(r)$ and ${{\mu }_{i}}\sim \mathcal{U}\{-1,1\}$ hold for all integer lattice sites $i\in \mathbb{Z}$, with $U_0 ​=0$. The transition probabilities ${{q}_{i,i-1}}$, ${{q}_{i,i+1}}$, and ${{q}_{i,i}}$ (denoting jumps to $i-1$, $i+1$, and remaining at $i$, respectively) depend on the random potential $U_i$, with strength and scale parameters $e$ and $c$.
	
	\begin{figure}[t]       
		\centering
		\includegraphics[width=0.9\linewidth]{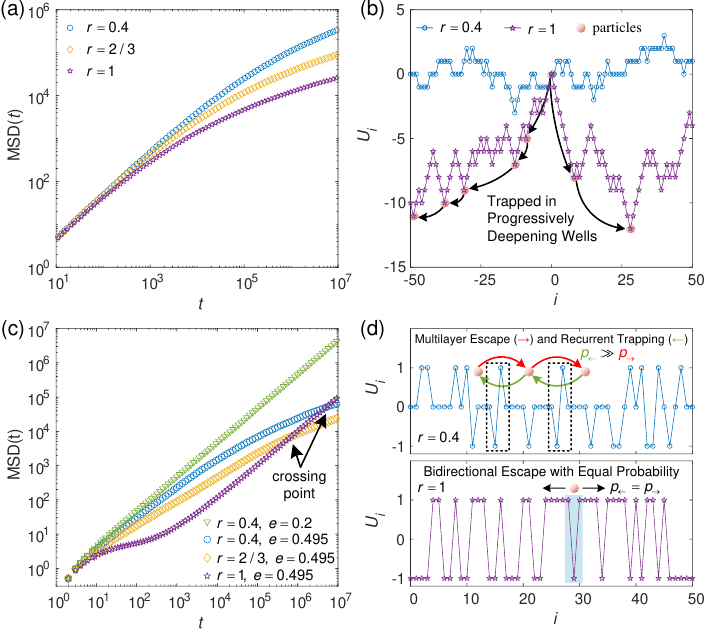}
		\subfigure{\label{fig:3a}}
		\subfigure{\label{fig:3b}}
		\subfigure{\label{fig:3c}}
		\subfigure{\label{fig:3d}}
		\caption{The ultraslow diffusion mechanism in our stochastic system $Q_2$ differs from the Sinai model. (a) The MSD of the Sinai model shows ultraslow diffusion for all $r$. This arises because (b) its potential variations follow a Brownian path. (c) For system $Q_2$, the MSD shows a parameter dependence largely consistent with that of map $Q_1$ (Fig.~\ref{fig:2}). This correspondence originates from (d) directional channels formed by symmetry breaking in local potential configurations within $0<r<1$ (boxed region, upper panel). Conversely, at $r=1$, these directional channels are globally eliminated (lower panel), and the system reverts to normal diffusion.}
		\label{fig:3}
	\end{figure}
	
	At $e=0.5$ and $c=2$, the MSD in the Sinai model exhibits a systematic decrease with increasing $r$ [Fig. \ref{fig:3a}, initial condition ${{i}_{0}} = 0$]. This observation stems from the thermally activated motion within the random potential landscape constructed from Brownian paths: Particles inherently tend toward lower energy states, progressively becoming trapped in deeper potential wells (probabilistic localization), with their mean exploration range growing as $({\ln t})^2$ \cite{bouchaud1990anomalous,bouchaud1990classical}. This mechanism aligns with physical intuition: The recurrent nature of Brownian paths ensures that zero-crossings of the potential field can occur at arbitrarily distant positions from the origin \cite{biskup2020return}. However, the distribution of equipotential points at $r<1$ significantly retards this process. The schematic potential profiles $U_i$​ in Fig. \ref{fig:3b} provide a visualization of this interpretation.
	
	In contrast to the classical Sinai model, our map $Q_1$ exhibits counterintuitive nonmonotonic transport behavior: on long timescales, the MSD in the fully disordered case $r=1$ eventually exceeds those in the sparse disorder cases $0<r<1$, as indicated by the MSD crossing point in Fig.~\ref{fig:2}. This suggests that the map $Q_1$ and Sinai dynamics are governed by different microscopic mechanisms. This distinction is also reflected in the potential profiles shown in the insets of Figs.~\ref{fig:1b}--\ref{fig:1d}: the random potential in the map $Q_1$ does not resemble a Brownian-path-like potential landscape, but is instead composed of independent and identically distributed (i.i.d.) local barrier and well configurations of uniform strength. Furthermore, extending thermally activated processes in random Brownian potential landscapes to deterministic dynamical systems typically requires Markov partitions as a fundamental premise~\cite{radons1996suppression}; by contrast, the diffusion process observed in this Letter does not rely on such structural constraints.
	
	To bridge this conceptual gap, we construct our second system $Q_2$ based on the Sinai model, where the random potential ${{{U}}_{i}}$ is defined as
	\begin{equation}
		\label{eq:5}
			{{U}_{i}}={{\xi }_{i}} {{\mu }_{i}},
	\end{equation}
	with ${{\xi }_{i}}$ and ${{\mu }_{i}}$ aligning with Eq. (\ref{eq:4}). Similar to the map $Q_1$​ (which ensures a finite-strength potential landscape), the system $Q_2​$ [Eq. (\ref{eq:3}) with Eq. (\ref{eq:5})] incorporates analogous constraints. Figure \ref{fig:3c} presents the MSD results for system $Q_2$ at $c=2$, showing the dependence of the MSD on $r$ and $e$ (with the critical value given by $e_{\mathrm c}=1/2$ for $c \geq 2$). A comparison with the results for map $Q_1$ in Fig. \ref{fig:2} shows that $Q_2$ exhibits a highly similar parameter dependence in its diffusion behavior.
	 
	The microscopic dynamics originate from symmetry breaking in local potential. At $r=1$, the system $Q_2$ exhibits symmetric ascent/descent profiles across all effective wells [lower panel of Fig. \ref{fig:3d}], ensuring equal bidirectional escape probabilities (denoted as ${{p}_{\leftarrow }}={{p}_{\to }}$). Although the system initially demonstrates transient ultraslow diffusion [with the timescale determined by the mean first-passage time $\left\langle {{T}^{*}} \right\rangle \approx \frac{3(2+c)}{4(1-2e)}$ required to cross the first barrier ($+1$), for Fig. \ref{fig:3c}, $\left\langle {{T}^{*}} \right\rangle \approx 300$], it eventually recovers to normal diffusion over extended evolution. Conversely, when $0<r<1$, local symmetry breaking characterized by “\systemN” and “\systemNN” shaped potential configurations emerges [an example is illustrated by the boxed region in the upper panel of Fig. \ref{fig:3d}, with detailed analysis provided in Sec. II of the SM]. When particles traverse such regions, the local escape probability in one direction becomes significantly dominant over the reverse process, as exemplified in the upper panel of Fig. \ref{fig:3d}, where leftward escape probability far exceeds the rightward counterpart (i.e., ${{p}_{\leftarrow }}\gg {{p}_{\to }}$). This effect reaches maximum prominence near critical strengths $\varepsilon _{\mathrm c} = 1/3$ (in the map $Q_1$) and $e _{\mathrm c}=1/2$ (in the system $Q_2$). We categorize these configurations as ​​directional channels​​: ​​\textit{reverse channels}​​ (severely restricted escape from one side) and ​​\textit{forward channels​​} (preferential diffusion).
	
	We now derive the logarithmic diffusion scaling from the coarse-grained directional-channel picture for sparse disorder near the critical potential strength. Since directional channels arise from quenched, i.i.d. well--barrier couplings, their spatial distribution is statistically translation-invariant; namely, a directional channel occurs with the same probability at any coarse-grained spatial position. Let $r_0$ denote the probability of finding a single reverse channel at a coarse-grained position. Since no global directional bias is imposed, the probability of finding a single forward channel is also $r_0$. However, within a finite spatial interval already explored by the particle, the numbers of forward and reverse channels exhibit a typical imbalance due to quenched disorder fluctuations. Let $L_x$ denote the typical number of excess reverse channels that controls the dominant escape time after the particle has explored a spatial scale $x$. According to the central limit theorem, the net imbalance between forward and reverse channels grows only as the square root of the explored distance, giving
	\begin{equation}
		\label{eq:6}
		L_x \sim (2r_0 x)^{1/2}.
	\end{equation}

	We next consider the time required to penetrate a single reverse channel. Near the critical potential strength, the occlusion effect of a reverse channel becomes strongly enhanced, so that the probability $p_{\rm r}$ for a particle to penetrate a single reverse channel within one microscopic time step is very small. Therefore, the characteristic penetration time scale of a single reverse channel can be defined as $\Delta t = p_{\rm r}^{-1} \gg 1$. After a particle penetrates a reverse channel, the channels that it has previously crossed become forward channels favoring the return direction, and may therefore recapture the particle into previously visited regions. Thus, the total escape process is not a simple one-way penetration process, but is controlled by capture--escape--recapture cycles. The recursion relation for this process and its asymptotic solution are given in Sec. III of the SM. The leading result is that the escape time required to overcome $L_x$ excess reverse channels grows exponentially with $L_x$:
	\begin{equation}
		\label{eq:7}
		T_{L_x} \sim \tau (\Delta t)^{L_x},
	\end{equation}
	where $\tau$ is a nonuniversal effective microscopic time scale. It absorbs the initial-level escape time, finite product terms, summation prefactors, and other constant time factors generated by the recapture process. These factors renormalize the time scale, but do not change the leading exponential dependence of $T_{L_x}$ on $L_x$.
	
	The typical spatial scale $x$ that the particle can explore within an observation time $t$ is determined by $t \sim T_{L_x}$. Combining Eqs. (\ref{eq:6}) and (\ref{eq:7}), we obtain ${t}/{\tau} \sim (\Delta t)^{(2r_0 x)^{1/2}}$. Taking the logarithm gives $\ln(t/\tau) \sim (2r_0 x)^{1/2}\ln (\Delta t)$. Thus, $x(t) \sim (2r_0)^{-1} \left[ {\ln(t/\tau)}/{\ln (\Delta t)} \right]^2$. In this hierarchical escape mechanism, the leading scaling of the MSD is set by the square of the typical explored distance, namely $\mathrm{MSD}(t) \sim x^2(t)$. We therefore obtain
	\begin{equation}
		\label{eq:8}
		\mathrm{MSD}(t) \sim A[\ln(t/\tau)]^4,
	\end{equation}
	where $A \propto r_0^{-2}[\ln (\Delta t)]^{-4}$ is a nonuniversal coefficient. Equation (\ref{eq:8}), as a scaling prediction of the coarse-grained directional-channel theory, shows that the anomalous diffusion observed in this Letter is not merely slowed diffusion resulting from a reduced ordinary diffusion coefficient or power-law subdiffusion. Instead, it is logarithmically ultraslow diffusion resulting from the combination of the exponential dependence of $T_{L_x}$ on $L_x$, generated by capture--escape--recapture dynamics, and the channel imbalance fluctuation $L_x\sim x^{1/2}$ in unbiased quenched disorder.
	 
	To directly test the logarithmic fourth-power scaling in Eq. (\ref{eq:8}), we further employ a fourth-root linear representation. If Eq. (\ref{eq:8}) holds, then $\mathrm{MSD}^{1/4}(t) \sim A^{1/4}\ln(t/\tau)$, or equivalently, $\mathrm{MSD}^{1/4}(t) \sim A^{1/4}\ln t - A^{1/4}\ln\tau$. Therefore, when $t$ is shown on a logarithmic horizontal axis, $\mathrm{MSD}^{1/4}(t)$ should become approximately linear in the long-time regime. Here the nonuniversal factor $A^{1/4}$ determines the slope, while the nonuniversal time scale $\tau$ is absorbed into the intercept.
	
	\begin{figure}[t]       
		\centering
		\includegraphics[width=0.9\linewidth]{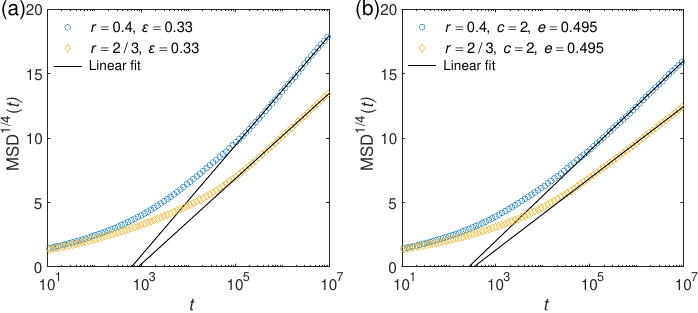}
		\subfigure{\label{fig:4a}}
		\subfigure{\label{fig:4b}}
		\caption{Fourth-root linearity test of the logarithmic fourth-power scaling. If Eq. (\ref{eq:8}) holds, then $\mathrm{MSD}^{1/4}(t)$ should be asymptotically linear in $\ln t$. In a representation with $t$ shown on a logarithmic horizontal axis, this relation appears as long-time linear behavior. (a) Test for the nonlinear map $Q_1$. (b) Test for the stochastic system $Q_2$. Colored symbols denote the numerical data for $\mathrm{MSD}^{1/4}(t)$, and black solid lines denote linear fits over the time window $t\in[1.0\times10^5,1.0\times10^7]$. All fits give coefficients of determination $R^2>0.999$.}
		\label{fig:4}
	\end{figure}

	Figures \ref{fig:4a} and \ref{fig:4b} show this test for the nonlinear map $Q_1$ and the stochastic system $Q_2$, respectively. For the ultraslow diffusion parameter sets, $\mathrm{MSD}^{1/4}(t)$ exhibits good linearity with $\ln t$ over the long-time window $t\in[1.0\times10^5,1.0\times10^7]$. All linear fits give coefficients of determination $R^2>0.999$. This consistency check supports the logarithmic fourth-power scaling prediction given by Eq. (\ref{eq:8}).
	
	In our first map $Q_1$, directional channels manifest as “\mapN” and “\mapNN” configurations [an example is shown within the boxed region in the inset of Fig. \ref{fig:1c}] due to the coupling between adjacent wells and barriers. The particles ​​become​​ trapped within a region enclosed by a pair of oppositely oriented directional channels (ochre-shaded regions). They frequently approach the boundary of the reverse channel and are reflected, causing them to spend most of their time oscillating back and forth within this confined area and thereby generating high-frequency fluctuations. Until an escape event occurs (red arrows), the particles will enter a larger-scale self-similar region (pale-yellow-shaded areas) composed of another pair of oppositely oriented directional channels. Crucially, recapture effects from preceding channels induce high-probability backtracking (green arrows). These recurrent capture-escape-recapture cycles persist across arbitrarily large spatial scales, progressively decelerating particle motion. Consequently, the diffusive spreading exhibits continuous slowdown, manifesting macroscopically as ultraslow diffusion in ensemble-averaged measurements. In the same near-critical sparse-disorder regime, the system further displays aging, breakdown of self-averaging, and weak ergodicity breaking (see Sec. IV of the SM).

	When $\varepsilon$ or $e$ deviates significantly from the critical value, bidirectional diffusion probabilities approach near-equilibrium [${{p}_{\leftarrow }}\approx {{p}_{\to }}$, see color-shaded regions in Fig. \ref{fig:1b} and the verification in Sec. II of the SM], despite persistent local weak symmetry breaking in directional channels. This permits the particles to effectively explore large regions of space within finite timeframes, which averages out the weak local symmetry breaking \cite{limoge1990temperature,mussawisade1997combination}. As shown in Figs. \ref{fig:2} and \ref{fig:3c}, normal diffusion is restored under these conditions on very long timescales (a conclusion further supported over a broader parameter range, as detailed in Sec. V of the SM). Remarkably, with appropriate parameters, the MSD of system $Q_2$ shows excellent agreement with that of map $Q_1$, as indicated by the black solid lines in Fig. \ref{fig:2}. This correspondence substantiates that the stochastic system $Q_2$ shares identical underlying mechanisms with our deterministic framework. To verify the generality of our results, we employed a Sine-lifted nonlinear map (Sec. VI) in the SM. Under these configurations, ultraslow diffusion remains observable. Furthermore, when the strength $\varepsilon$ exhibits fluctuations (Sec. VII), this anomalous dynamics persists, indicating that the fundamental mechanism we propose for generating ultraslow diffusion is highly robust.

	In summary, we have demonstrated that sparsely distributed wells and barriers mediate qualitative changes in diffusion dynamics via local symmetry breaking. This may explain a long-standing puzzle: why certain disordered systems exhibit stronger suppression of diffusion at lower disorder densities than at higher ones \cite{havlin1987diffusion}. Let us return to the first example \cite{song2010modelling}. When a journey concludes, you tend to return to your origin, with long-range displacements only resuming at the next journey (this is not a frequent occurrence for the majority). Such movement patterns remain “anchored” near core locations like homes and workplaces [analogous to the color-shaded regions in Fig. \ref{fig:1c}], exhibiting repetitive visitation (green arrows) while demonstrating infrequent and hierarchical exploration of new areas (red arrows). The underlying mechanism remarkably parallels our findings, suggesting this Letter may offer an alternative explanatory perspective for this well-known social phenomenon. Another noteworthy aspect is that previous studies have established the Sinai mechanism's fundamental reliance on potential long-range correlations, regardless of dimensionality \cite{goychuk2017persistent,fisher1984random,stein1995broken,selinger1989diffusion}. In contrast, our work breaks with this paradigm, as the wells and barriers in our model are i.i.d. To what extent this mechanism generates ultraslow diffusion in higher-dimensional scenarios presents an intriguing and important open question.

	\vspace{\baselineskip}
	This work was supported by the National Natural Science Foundation of China (Grants No. 62471496, No. 62071496 and No. 62671629) and the Research and Innovation Project of Graduate of Central South University, China (Project No. 2025ZZTS0610). We are grateful to Dr. Yao Chen for insightful discussions. We also thank the High-Performance Computing Center at Central South University for providing essential computational resources.
	
	\vspace{\baselineskip}
	
	\textit{Data availability.} There are no publicly available research data or software supporting this manuscript. Requests for further information or data should be sent to the authors.
	\setcitestyle{numbers}
	
	\bibliographystyle{apsrev4-2}
	\nocite{*}
	\bibliography{mybibfile}

@article{vaknin2000aging,
	title={Aging Effects in an Anderson Insulator},
	author={Vaknin, A and Ovadyahu, Zvi and Pollak, M},
	journal={Phys. Rev. Lett.},
	volume={84},
	number={15},
	pages={3402},
	year={2000},
	publisher={APS},
	url = {https://doi.org/10.1103/PhysRevLett.84.3402}
}

@article{brauns2002complex,
	title={Complex Local Dynamics in DNA on the Picosecond and Nanosecond Time Scales},
	author={Brauns, Eric B and Madaras, Mihaela L and Coleman, Robert S and Murphy, Catherine J and Berg, Mark A},
	journal={Phys. Rev. Lett.},
	volume={88},
	number={15},
	pages={158101},
	year={2002},
	publisher={APS},
	url = {https://doi.org/10.1103/PhysRevLett.88.158101}
}

@article{matan2002crumpling,
	title={Crumpling a Thin Sheet},
	author={Matan, Kittiwit and Williams, Rachel B and Witten, Thomas A and Nagel, Sidney R},
	journal={Phys. Rev. Lett.},
	volume={88},
	number={7},
	pages={076101},
	year={2002},
	publisher={APS},
	url = {https://doi.org/10.1103/PhysRevLett.88.076101}
}

@article{bendavid2010slipstick,
	title={Slip-stick and the evolution of frictional strength},
	author={Ben-David, Oded and Rubinstein, Shmuel M and Fineberg, Jay},
	journal={Nature},
	volume={463},
	number={7277},
	pages={76--79},
	year={2010},
	publisher={Nature Publishing Group UK London},
	url = {https://doi.org/10.1038/nature08676}
}

@article{richard2005slow,
	title={Slow relaxation and compaction of granular systems},
	author={Richard, Patrick and Nicodemi, Mario and Delannay, Renaud and Ribiere, Philippe and Bideau, Daniel},
	journal={Nat. Mater.},
	volume={4},
	number={2},
	pages={121--128},
	year={2005},
	publisher={Nature Publishing Group UK London},
	url = {https://doi.org/10.1038/nmat1300}
}

@article{boettcher2011ageing,
	title={Ageing in dense colloids as diffusion in the logarithm of time},
	author={Boettcher, Stefan and Sibani, Paolo},
	journal={J. Phys.: Condens. Matter},
	volume={23},
	number={6},
	pages={065103},
	year={2011},
	publisher={IOP Publishing},
	url = {https://doi.org/10.1088/0953-8984/23/6/065103}
}

@article{watanabe2018empirical,
	title={Empirical observations of ultraslow diffusion driven by the fractional dynamics in languages},
	author={Watanabe, Hayafumi},
	journal={Phys. Rev. E},
	volume={98},
	number={1},
	pages={012308},
	year={2018},
	publisher={APS},
	url = {https://doi.org/10.1103/PhysRevE.98.012308}
}

@article{eab2011fractional,
	title={Fractional Langevin equations of distributed order},
	author={Eab, CH and Lim, SC},
	journal={Phys. Rev. E},
	volume={83},
	number={3},
	pages={031136},
	year={2011},
	publisher={APS},
	url = {https://doi.org/10.1103/PhysRevE.83.031136}
}

@article{kochubei2008distributed,
	title={Distributed order calculus and equations of ultraslow diffusion},
	author={Kochubei, Anatoly N},
	journal={J. Math. Anal. Appl.},
	volume={340},
	number={1},
	pages={252--281},
	year={2008},
	publisher={Elsevier},
	url = {https://doi.org/10.1016/j.jmaa.2007.08.024}
}

@article{liang2018nonlocal,
	title={A non-local structural derivative model for characterization of ultraslow diffusion in dense colloids},
	author={Liang, Yingjie and Chen, Wen},
	journal={Commun. Nonlinear Sci. Numer. Simul.},
	volume={56},
	pages={131--137},
	year={2018},
	publisher={Elsevier},
	url = {https://doi.org/10.1016/j.cnsns.2017.07.027}
}

@article{liang2018diffusion,
	title={Diffusion entropy method for ultraslow diffusion using inverse Mittag-Leffler function},
	author={Liang, Yingjie},
	journal={Fract. Calc. Appl. Anal.},
	volume={21},
	number={1},
	pages={104--117},
	year={2018},
	publisher={De Gruyter},
	url = {https://doi.org/10.1515/fca-2018-0007}
}

@article{sandev2016comb,
	title={Comb Model with Slow and Ultraslow Diffusion},
	author={Sandev, Trifce and Iomin, Alexander and Kantz, Holger and Metzler, Ralf and Chechkin, Aleksei},
	journal={Math. Model. Nat. Phenom.},
	volume={11},
	number={3},
	pages={18--33},
	year={2016},
	publisher={EDP Sciences},
	url = {https://doi.org/10.1051/mmnp/201611302}
}

@article{havlin1987anomalous,
	title={Anomalous diffusion on a random comblike structure},
	author={Havlin, Shlomo and Kiefer, James E and Weiss, George H},
	journal={Phys. Rev. A},
	volume={36},
	number={3},
	pages={1403},
	year={1987},
	publisher={APS},
	url = {https://doi.org/10.1103/PhysRevA.36.1403}
}

@article{iomin2016ultraslow,
	title={Does ultra-slow diffusion survive in a three dimensional cylindrical comb?},
	author={Iomin, A and M{\'e}ndez, V},
	journal={Chaos Solitons Fract.},
	volume={82},
	pages={142--147},
	year={2016},
	publisher={Elsevier},
	url = {https://doi.org/10.1016/j.chaos.2015.11.017}
}

@article{havlin1990longtailed,
	title={A new class of long-tailed pausing time densities for the CTRW},
	author={Havlin, Shlomo and Weiss, George H},
	journal={J. Stat. Phys.},
	volume={58},
	pages={1267--1273},
	year={1990},
	publisher={Springer},
	url = {https://link.springer.com/article/10.1007/BF01026577}
}

@article{meerschaert2006stochastic,
	title={Stochastic model for ultraslow diffusion},
	author={Meerschaert, Mark M and Scheffler, Hans-Peter},
	journal={Stoch. Process. Their Appl.},
	volume={116},
	number={9},
	pages={1215--1235},
	year={2006},
	publisher={Elsevier},
	url = {https://doi.org/10.1016/j.spa.2006.01.006}
}

@article{denisov2013limiting,
	title={Limiting distributions of continuous-time random walks with superheavy-tailed waiting times},
	author={Denisov, SI and Bystrik, Yu S and Kantz, H},
	journal={Phys. Rev. E},
	volume={87},
	number={2},
	pages={022117},
	year={2013},
	publisher={APS},
	url = {https://doi.org/10.1103/PhysRevE.87.022117}
}

@article{denisov2011asymptotic,
	title={Asymptotic solutions of decoupled continuous-time random walks with superheavy-tailed waiting time and heavy-tailed jump length distributions},
	author={Denisov, SI and Yuste, SB and Bystrik, Yu S and Kantz, H and Lindenberg, K},
	journal={Phys. Rev. E},
	volume={84},
	number={6},
	pages={061143},
	year={2011},
	publisher={APS},
	url = {https://doi.org/10.1103/PhysRevE.84.061143}
}

@article{radons1996suppression,
	title={Suppression of Chaotic Diffusion by Quenched Disorder},
	author={Radons, G{\"u}nter},
	journal={Phys. Rev. Lett.},
	volume={77},
	number={23},
	pages={4748},
	year={1996},
	publisher={APS},
	url = {https://doi.org/10.1103/PhysRevLett.77.4748}
}

@article{radons2004anomalous,
	title={Anomalous transport in disordered dynamical systems},
	author={Radons, G{\"u}nter},
	journal={Physica D},
	volume={187},
	number={1-4},
	pages={3--19},
	year={2004},
	publisher={Elsevier},
	url = {https://doi.org/10.1016/j.physd.2003.09.001}
}

@article{barkai2003aging,
	title={Aging in Subdiffusion Generated by a Deterministic Dynamical System},
	author={Eli Barkai},
	journal={Phys. Rev. Lett.},
	year={2003},
	volume={90},
	pages={104101},
	url={https://doi.org/10.1103/PHYSREVLETT.90.104101}
}

@article{geisel1982onset,
	title={Onset of Diffusion and Universal Scaling in Chaotic Systems},
	author={Theo Geisel and J. Nierwetberg},
	journal={Phys. Rev. Lett.},
	year={1982},
	volume={48},
	pages={7-10},
	url={https://doi.org/10.1103/PHYSREVLETT.48.7}
}

@article{fujisaka1982chaos,
	title={Chaos-induced diffusion in nonlinear discrete dynamics},
	author={Fujisaka, H and Grossmann, S},
	journal={Z. Phys. B Condens. Matter},
	volume={48},
	pages={261--275},
	year={1982},
	publisher={Springer},
	url = {https://doi.org/10.1007/BF01420589}
}

@article{schell1982diffusive,
	title={Diffusive dynamics in systems with translational symmetry: A one-dimensional-map model},
	author={Schell, Mark and Fraser, Simon and Kapral, Raymond},
	journal={Phys. Rev. A},
	volume={26},
	number={1},
	pages={504-521},
	year={1982},
	publisher={APS},
	url = {https://doi.org/10.1103/PhysRevA.26.504}
}

@article{gaspard1998chaotic,
	title={Chaotic and fractal properties of deterministic diffusion-reaction processes},
	author={Gaspard, Pierre and Klages, R},
	journal={Chaos},
	volume={8},
	number={2},
	pages={409--423},
	year={1998},
	publisher={American Institute of Physics},
	url = {https://doi.org/10.1063/1.166323}
}

@article{klages1995simple,
	title={Simple Maps with Fractal Diffusion Coefficients},
	author={Klages, R and Dorfman, JR},
	journal={Phys. Rev. Lett.},
	volume={74},
	number={3},
	pages={387},
	year={1995},
	publisher={APS},
	url = {https://doi.org/10.1103/PhysRevLett.74.387}
}

@article{klages1999simple,
	title={Simple deterministic dynamical systems with fractal diffusion coefficients},
	author={Klages, R and Dorfman, JR},
	journal={Phys. Rev. E},
	volume={59},
	number={5},
	pages={5361},
	year={1999},
	publisher={APS},
	url = {https://doi.org/10.1103/PhysRevE.59.5361}
}

@article{sarkar2023quench,
	title={Quench-induced chaotic dynamics of Anderson-localized interacting Bose-Einstein condensates in one dimension},
	author={Sarkar, Swarup K and Mishra, Tapan and Muruganandam, Paulsamy and Mishra, Pankaj K},
	journal={Phys. Rev. A},
	volume={107},
	number={5},
	pages={053320},
	year={2023},
	publisher={APS},
	url = {https://doi.org/10.1103/PhysRevA.107.053320}
}

@article{liu2024quenched,
	title={Quenched disorder-induced chaotic resonance: Optimal stimulus response near the edge of chaos},
	author={Liu, Cong and Wu, Zhi-Xi and Guan, Jian-Yue},
	journal={Phys. Rev. Res.},
	volume={6},
	number={4},
	pages={L042019},
	year={2024},
	publisher={APS},
	url = {https://doi.org/10.1103/PhysRevResearch.6.L042019}
}

@article{sato2019anomalous,
	title={Anomalous Diffusion in Random Dynamical Systems},
	author={Sato, Yuzuru and Klages, Rainer},
	journal={Phys. Rev. Lett.},
	volume={122},
	number={17},
	pages={174101},
	year={2019},
	publisher={APS},
	url = {https://doi.org/10.1103/PhysRevLett.122.174101}
}

@book{klages2007microscopic,
	title={Microscopic chaos, fractals and transport in nonequilibrium statistical mechanics},
	author={Klages, Rainer},
	volume={24},
	year={2007},
	publisher={World Scientific},
}

@article{felderhof2008escape,
	title={Escape by diffusion from a square well across a square barrier},
	author={Felderhof, BU},
	journal={Physica A},
	volume={387},
	number={1},
	pages={39--56},
	year={2008},
	publisher={Elsevier},
	url = {https://doi.org/10.1016/j.physa.2007.08.041}
}

@misc{supplementalMaterial,
	note={See Supplemental Material at \url{https://link.aps.org/supplemental/10.1103/cvmv-86bh} for additional analytical and numerical results, which includes Refs. \cite{klafter2011first,norris1998markov,lomholt2013microscopic,li2025route,dentz2016selfaveraging,huberman1980noise,dhumieres1982chaotic}}
}

@book{norris1998markov,
	title={Markov chains},
	author={Norris, James R},
	number={2},
	year={1998},
	publisher={Cambridge university press}
}

@article{lomholt2013microscopic,
	title={Microscopic Origin of the Logarithmic Time Evolution of Aging Processes in Complex Systems},
	author={Lomholt, Michael A and Lizana, Ludvig and Metzler, Ralf and Ambj{\"o}rnsson, Tobias},
	journal={Phys. Rev. Lett.},
	volume={110},
	number={20},
	pages={208301},
	year={2013},
	publisher={APS},
	url = {https://doi.org/10.1103/PhysRevLett.110.208301}
}

@article{li2025route,
	title={Route of Random Process to Ultraslow Aging Phenomena},
	author={Li, Chunyan and Liu, Haiwen and Xie, XC},
	journal={Phys. Rev. Lett.},
	volume={134},
	number={19},
	pages={197102},
	year={2025},
	publisher={APS},
	url = {https://doi.org/10.1103/PhysRevLett.134.197102}
}

@article{dentz2016selfaveraging,
	title={Self-averaging and ergodicity of subdiffusion in quenched random media},
	author={Dentz, Marco and Russian, Anna and Gouze, Philippe},
	journal={Phys. Rev. E},
	volume={93},
	number={1},
	pages={010101},
	year={2016},
	publisher={APS},
	url = {https://doi.org/10.1103/PhysRevE.93.010101}
}

@article{huberman1980noise,
	title={Noise phenomena in Josephson junctions},
	author={Bernardo A. Huberman and James P. Crutchfield and Norman H. Packard},
	journal={Appl. Phys. Lett.},
	year={1980},
	volume={37},
	pages={750-752},
	url={https://doi.org/10.1063/1.92020}
}

@article{dhumieres1982chaotic,
	title={Chaotic states and routes to chaos in the forced pendulum},
	author={Dominique d'Humi{\`e}res and Malcolm R. Beasley and Bernardo A. Huberman and Albert Libchaber},
	journal={Phys. Rev. A},
	year={1982},
	volume={26},
	pages={3483-3496},
	url={https://doi.org/10.1103/PHYSREVA.26.3483}
}

@book{klafter2011first,
	title={First steps in random walks: from tools to applications},
	author={Klafter, Joseph and Sokolov, Igor M},
	year={2011},
	publisher={OUP Oxford},
	url = {https://www.google.com/books?hl=en&lr=&id=zxVREAAAQBAJ&oi=fnd&pg=PP1&dq=First+steps+in+random+walks:+from+tools+to+applications&ots=8XACpsh0YD&sig=ObD2vwre9Srj_6FsLyg9YFrB7GU}
}

@article{dasilva2019correlated,
	title={Correlated Brownian motion and diffusion of defects in spatially extended chaotic systems},
	author={da Silva, ST and Prado, T de L and Lopes, SR and Viana, RL},
	journal={Chaos},
	volume={29},
	number={7},
	year={2019},
	pages={071104},
	publisher={AIP Publishing},
	url = {https://doi.org/10.1063/1.5113783}
}

@article{chave1999statistical,
	title={Statistical and dynamical properties of the discrete Sinai model at finite times},
	author={Chave, Jerome and Guitter, Emmanuel},
	journal={J. Phys. A: Math. Gen.},
	volume={32},
	number={3},
	pages={445},
	year={1999},
	publisher={IOP Publishing},
	url = {https://doi.org/10.1088/0305-4470/32/3/003}
}

@article{padash2022local,
	title={Local equilibrium properties of ultraslow diffusion in the Sinai model},
	author={Padash, Amin and Aghion, Erez and Schulz, Alexander and Barkai, Eli and Chechkin, Aleksei V and Metzler, Ralf and Kantz, Holger},
	journal={New J. Phys.},
	volume={24},
	number={7},
	pages={073026},
	year={2022},
	publisher={IOP Publishing},
	url = {https://doi.org/10.1088/1367-2630/ac7df8}
}

@article{biskup2020return,
	title={Return Probability and Recurrence for the Random Walk Driven by Two-Dimensional Gaussian Free Field},
	author={Biskup, Marek and Ding, Jian and Goswami, Subhajit},
	journal={Commun. Math. Phys.},
	volume={373},
	number={1},
	pages={45--106},
	year={2020},
	publisher={Springer},
	url = {https://doi.org/10.1007/s00220-019-03589-z}
}

@article{bouchaud1990classical,
	title={Classical diffusion of a particle in a one-dimensional random force field},
	author={Bouchaud, J-Ph and Comtet, A and Georges, A and Le Doussal, P},
	journal={Ann. Phys.},
	volume={201},
	number={2},
	pages={285--341},
	year={1990},
	publisher={Elsevier},
	url = {https://doi.org/10.1016/0003-4916(90)90043-N}
}

@article{goychuk2017persistent,
	title={Persistent Sinai-type diffusion in Gaussian random potentials with decaying spatial correlations},
	author={Goychuk, Igor and Kharchenko, Vasyl O and Metzler, Ralf},
	journal={Phys. Rev. E},
	volume={96},
	number={5},
	pages={052134},
	year={2017},
	publisher={APS},
	url = {https://doi.org/10.1103/PhysRevE.96.052134}
}

@article{geisel1984anomalous,
	title={Anomalous Diffusion in Intermittent Chaotic Systems},
	author={Geisel, Theo and Thomae, Stefan},
	journal={Phys. Rev. Lett.},
	volume={52},
	number={22},
	pages={1936},
	year={1984},
	publisher={APS},
	url = {https://doi.org/10.1103/PhysRevLett.52.1936}
}

@article{geisel1985accelerated,
	title={Accelerated Diffusion in Josephson Junctions and Related Chaotic Systems},
	author={Geisel, T and Nierwetberg, J and Zacherl, A},
	journal={Phys. Rev. Lett.},
	volume={54},
	number={7},
	pages={616},
	year={1985},
	publisher={APS},
	url = {https://doi.org/10.1103/PhysRevLett.54.616}
}

@article{zumofen1993scaleinvariant,
	title={Scale-invariant motion in intermittent chaotic systems},
	author={Zumofen, G and Klafter, J},
	journal={Phys. Rev. E},
	volume={47},
	number={2},
	pages={851},
	year={1993},
	publisher={APS},
	url = {https://doi.org/10.1103/PhysRevE.47.851}
}

@article{drager2000strong,
	title={Strong anomaly in diffusion generated by iterated maps},
	author={Dr{\"a}ger, Julia and Klafter, Joseph},
	journal={Phys. Rev. Lett.},
	volume={84},
	number={26},
	pages={5998},
	year={2000},
	publisher={APS},
	url = {https://doi.org/10.1103/PhysRevLett.84.5998}
}

@article{golosov1984localization,
	title={Localization of random walks in one-dimensional random environments},
	author={Golosov, AO},
	journal={Commun. Math. Phys.​},
	volume={92},
	pages={491--506},
	year={1984},
	publisher={Springer},
	url = {https://link.springer.com/article/10.1007/BF01215280}
}

@article{bouchaud1990anomalous,
	title={Anomalous diffusion in disordered media: statistical mechanisms, models and physical applications},
	author={Bouchaud, Jean-Philippe and Georges, Antoine},
	journal={Phys. Rep.​​},
	volume={195},
	number={4-5},
	pages={127--293},
	year={1990},
	publisher={Elsevier},
	url = {https://doi.org/10.1016/0370-1573(90)90099-N}
}

@article{barkai1997crossover,
	title={Crossover from dispersive to regular transport in biased maps},
	author={Barkai, E and Klafter, J},
	journal={Phys. Rev. Lett.},
	volume={79},
	number={12},
	pages={2245},
	year={1997},
	publisher={APS},
	url = {https://doi.org/10.1103/PhysRevLett.79.2245}
}

@article{klafter1996beyond,
	title={Beyond brownian motion},
	author={Klafter, Joseph and Shlesinger, Michael F and Zumofen, Gert},
	journal={Phys. Today},
	volume={49},
	number={2},
	pages={33--39},
	year={1996},
	publisher={American Institute of Physics},
	url = {https://doi.org/10.1063/1.881487}
}

@article{klages1997dynamical,
	title={Dynamical crossover in deterministic diffusion},
	author={Klages, R and Dorfman, JR},
	journal={Phys. Rev. E},
	volume={55},
	number={2},
	pages={R1247},
	year={1997},
	publisher={APS},
	url = {https://doi.org/10.1103/PhysRevE.55.R1247}
}

@article{popescu1998quenched,
	title={Quenched disorder enhances chaotic diffusion},
	author={Popescu, MN and Braiman, Y and Family, F and Hentschel, HGE},
	journal={Phys. Rev. E},
	volume={58},
	number={4},
	pages={R4057},
	year={1998},
	publisher={APS},
	url = {https://doi.org/10.1103/PhysRevE.58.R4057}
}

@article{korabel2007fractal,
	title={Fractal properties of anomalous diffusion in intermittent maps},
	author={Korabel, Nickolay and Klages, Rainer and Chechkin, Aleksei V and Sokolov, Igor M and Gonchar, Vsevolod Yu},
	journal={Phys. Rev. E},
	volume={75},
	number={3},
	pages={036213},
	year={2007},
	publisher={APS},
	url = {https://doi.org/10.1103/PhysRevE.75.036213}
}

@article{sinai1982limit,
	title={The Limiting Behavior of a One-Dimensional Random Walk in a Random Medium},
	author={Sinai, Yakov Grigor'evich},
	journal={Theory Probab. Appl.},
	volume={27},
	number={2},
	pages={247--258},
	year={1982},
	publisher={Russian Academy of Sciences, Steklov Mathematical Institute of Russian~…},
	url = {https://doi.org/10.1137/1127028}
}

@article{song2010modelling,
	title={Modelling the scaling properties of human mobility},
	author={Song, Chaoming and Koren, Tal and Wang, Pu and Barab{\'a}si, Albert-L{\'a}szl{\'o}},
	journal={Nat. Phys.},
	volume={6},
	number={10},
	pages={818--823},
	year={2010},
	publisher={Nature Publishing Group UK London},
	url = {https://doi.org/10.1038/NPHYS1760}
}

@article{havlin1987diffusion,
	title={Diffusion in disordered media},
	author={Shlomo, Havlin and Daniel, Ben-Avraham},
	journal={​​Adv. Phys.​​},
	volume={36},
	number={6},
	year={1987},
	pages={695--798},
	url = {https://doi.org/10.1080/00018738700101072}
}

@article{fisher1984random,
	title={Random walks in random environments},
	author={Fisher, Daniel S},
	journal={Phys. Rev. A},
	volume={30},
	number={2},
	pages={960},
	year={1984},
	publisher={APS},
	url = {https://doi.org/10.1103/PhysRevA.30.960}
}

@article{stein1995broken,
	title={Broken ergodicity and the geometry of rugged landscapes},
	author={Stein, Daniel L and Newman, Charles M},
	journal={Phys. Rev. E},
	volume={51},
	number={6},
	pages={5228},
	year={1995},
	publisher={APS},
	url = {https://doi.org/10.1103/PhysRevE.51.5228}
}

@article{selinger1989diffusion,
	title={Diffusion in the presence of quenched random bias fields: A two-dimensional generalization of the Sinai model},
	author={Selinger, Robin L Blumberg and Havlin, Shlomo and Leyvraz, Fran{\c{c}}ois and Schwartz, Moshe and Stanley, H Eugene},
	journal={Phys. Rev. A},
	volume={40},
	number={11},
	pages={6755},
	year={1989},
	publisher={APS},
	url = {https://doi.org/10.1103/PhysRevA.40.6755}
}

@article{degennes1975brownian,
	title={Brownian motion of a classical particle through potential barriers. Application to the helix-coil transitions of heteropolymers},
	author={De Gennes, PG},
	journal={​​J. Stat. Phys.​​},
	volume={12},
	number={6},
	pages={463--481},
	year={1975},
	publisher={Springer},
	url = {https://doi.org/10.1007/BF01025834}
}

@article{bassler1987viscous,
	title={Viscous flow in supercooled liquids analyzed in terms of transport theory for random media with energetic disorder},
	author={B{\"a}ssler, Heinz},
	journal={Phys. Rev. Lett.},
	volume={58},
	number={8},
	pages={767},
	year={1987},
	publisher={APS},
	url = {https://doi.org/10.1103/PhysRevLett.58.767}
}

@article{zwanzig1988diffusion,
	title={Diffusion in a rough potential},
	author={Zwanzig, Robert},
	journal={Proc. Natl. Acad. Sci. U.S.A.​​},
	volume={85},
	number={7},
	pages={2029--2030},
	year={1988},
	url = {https://doi.org/10.1073/pnas.85.7.2029}
}

@article{limoge1990temperature,
	title={Temperature behavior of tracer diffusion in amorphous materials: A random-walk approach},
	author={Limoge, Y and Bocquet, JL},
	journal={Phys. Rev. Lett.},
	volume={65},
	number={1},
	pages={60},
	year={1990},
	publisher={APS},
	url = {https://doi.org/10.1103/PhysRevLett.65.60}
}

@article{mussawisade1997combination,
	title={Combination of random-barrier and random-trap models},
	author={Mussawisade, K and Wichmann, T and Kehr, KW},
	journal={J. Phys.: Condens. Matter},
	volume={9},
	number={6},
	pages={1181},
	year={1997},
	publisher={IOP Publishing},
	url = {https://doi.org/10.1088/0953-8984/9/6/005}
}

@article{klages2002suppression,
	title={Suppression and enhancement of diffusion in disordered dynamical systems},
	author={Klages, R},
	journal={Phys. Rev. E},
	volume={65},
	number={5},
	pages={055203},
	year={2002},
	publisher={APS},
	url = {https://doi.org/10.1103/PhysRevE.65.055203}
}
\end{document}